\documentclass[amsmath,amssymb,aip,jap,reprint]{revtex4-1}
\usepackage{graphicx}% Include figure files
\usepackage{dcolumn}% Align table columns on decimal point
\usepackage{subfigure}
\usepackage{bm}% bold math
\usepackage{xcolor}

\begin{document}

%\preprint{APS/123-QED}

\title{Carbon Vacancy Formation in Binary Transition Metal Carbides from Density Functional Theory}

\author{Mikael R{\aa}sander}\email{mikael.rasander@ltu.se}
\affiliation{%
Applied Physics, Division of Materials Science, Department of Engineering Sciences and Mathematics, Lule{\aa} University of Technology, 971 87 Lule{\aa}, Sweden}
\author{Anna Delin}
\affiliation{%
Department of Applied Physics, School of Engineering Sciences, KTH Royal Institute of Technology, Electrum 229, 164 40 Kista, Sweden}
\affiliation{%
Department of Physics and Astronomy, Uppsala University, Box 516, 751 20 Uppsala, Sweden}
\affiliation{%
SeRC (Swedish e-Science Research Center), KTH Royal Institute of Technology, 100 44 Stockholm, Sweden}
%\affiliation{%
%Department of Materials Science and Engineering, KTH Royal Institute of Technology, Sweden
%}%

\date{\today}% It is always \today, today,

\begin{abstract}
We have investigated the trends in the formation of carbon vacancies in binary transition metal (TM) carbides using density functional calculations for two common TM carbide crystal structures, namely the B1 (rock-salt) and WC structure types. The TM are taken from group IV (Ti, Zr, Hf), V (V, Nb, Ta) and VI (Cr, Mo, W) of the periodic table, as well as Sc from group III. For B1-structured TM carbides, the general trend is that it is easier for C vacancies to be formed as the number of valence electrons in the system increases. The exception is ScC where C vacancies are rather easy to form. It is also clear that the formation of C vacancies depends on the growth conditions: For TM-rich conditions, B1-structured carbides will always favour C vacancy formation. For C-rich conditions it is energetically favourable to form C vacancies in ScC, VC, NbC, CrC, MoC and WC. Experimentally large C vacancy concentrations are found in B1-structured TM carbides and our calculations are in line with these observations. In fact, TiC, ZrC, HfC and TaC are the only B1-structured TM carbides that do not spontaneously favour C vacancies and then only for C-rich conditions. C vacancies may still be present in these systems, however, due to the high temperatures used when growing TM carbides. In the case of WC-structured TM carbides, C vacancy formation is not energetically favourable irrespective of the growth conditions, which is the reason why this structure is only found close to a one-to-one metal-to-carbon ratio. In addition, we have investigated the change in the local structure induced by the presence of vacancies as well as the associated relaxation energy. We find that relaxations of the atoms close to the C vacancy is smaller in the WC-structured carbides than in B1-structured carbides. The smaller relaxations in WC-structured carbides works towards a lower tendency for C vacancies to be formed in the WC-structured than in B1-structured carbides since relaxations always work towards stabilising the vacancy.
\end{abstract}

%\pacs{}% PACS, the Physics and Astronomy
\maketitle

\section{Introduction}
The binary transition metal (TM) carbides have been of interest for a long time due to their many interesting physical properties, such as high hardness, high melting temperatures, good thermal and electrical conductivities, and even superconductivity.\cite{toth,Schwarz1987,Williams1966,Williams1971,Williams1988}  These compounds are therefore of interest in many applications, for example as hard metal coatings,\cite{toth} in catalysis,\cite{Vojvodic2009} and as electrical contacts.\cite{Oberg2010} TM carbides can also be used in low-friction coatings\cite{Wilhelmsson,Jansson2013,Bijelovic2010,Rasander,Jansson2011} where the TM carbide phase is enclosed inside a matrix of amorphous carbon.
\par
\begin{figure*}[t]
\includegraphics[width=14cm]{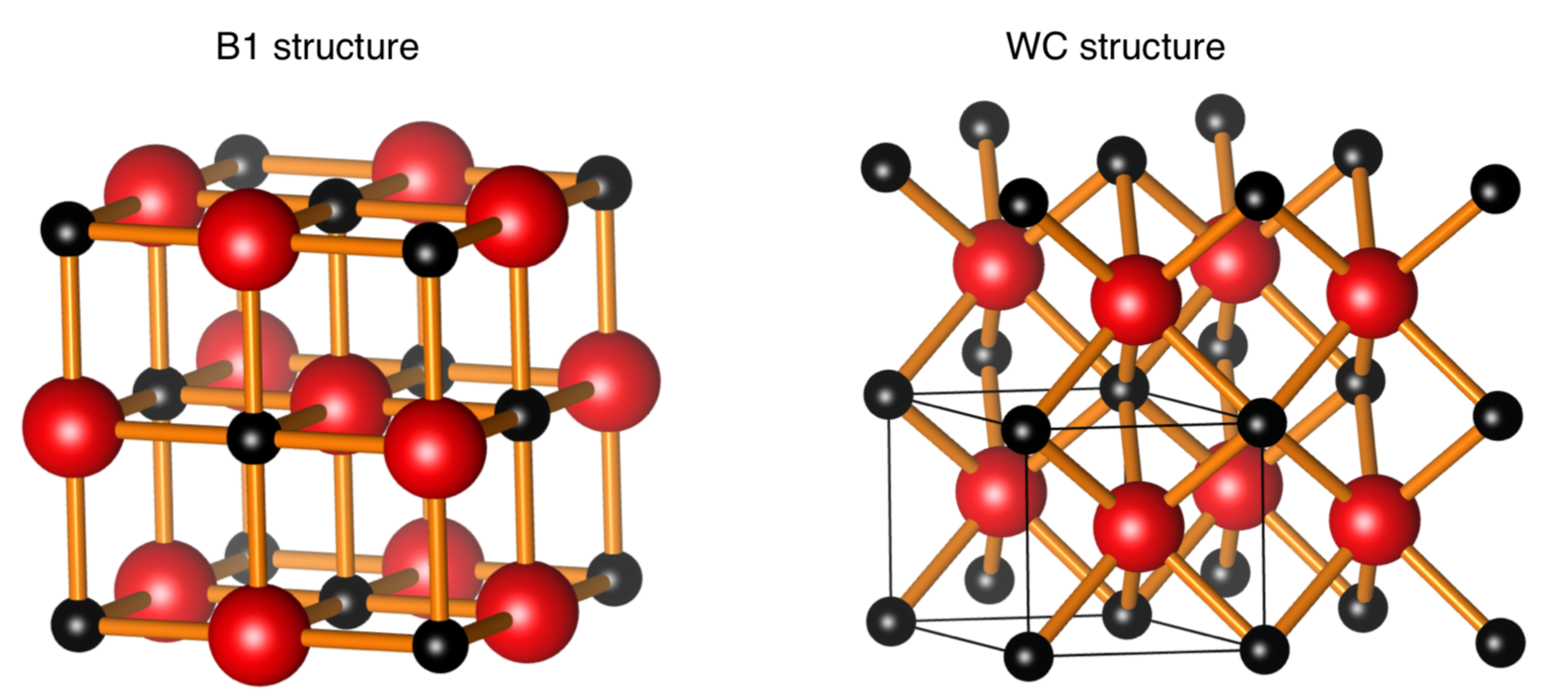}
\caption{\label{fig:structures} (Color online) Illustrations of the B1 and WC structures. The conventional B1 unit cell (left) and a part of a WC-structured TM carbide with the primitive cell outlined (right) are shown with red spheres representing TM atoms and black spheres representing carbon. The nearest neighbour bonds between TM and carbon atoms are also shown.}
\end{figure*}

Carbide formation is quite common among the TM; the strongest carbide forming metals are TM from Groups IV, V and VI. These systems form binary carbide phases with a more or less one-to-one ratio between TM and carbon.\cite{toth,Schwarz1987,Williams1966} Moving further along the periodic table carbide formation becomes weaker and ordered carbides with a much lower carbon content is formed only for the 3d TM, e.g. Cr$_{3}$C$_{2}$, Cr$_{7}$C$_{3}$ and Fe$_{3}$C, while the 4d and 5d TM beyond Tc and Re, respectively, only form carbides under extreme conditions.\cite{Jansson2013} Sc and Y also form TM carbides; while Sc can form simple carbides in the rock-salt crystal structure, Y forms carbides with more complex crystal structures.\cite{Jansson2013} Depending on the TM the phase diagrams of the TM carbides can be both rather simple and showing only a few different phases or very complex and displaying a large number of different phases. A discussion of the full phase diagrams of these compounds is, however, beyond the scope of the present paper. Even so, we note that the most common crystal structure in which TM carbides are found is the B1 or rock salt crystal structure, illustrated in Fig.~\ref{fig:structures}. This structure is found for ScC, TiC, VC, ZrC, NbC, MoC, HfC and TaC.\cite{toth,Jansson2013} Another common carbide structure type is the WC-structure, also illustrated in Fig.~\ref{fig:structures}, which is the archetypical hexagonal structure found for WC.\cite{toth,Hugosson2003} This structure is also the ground state of MoC.\cite{Hugosson-1999,Hugosson-2001b,Hugosson-2001c} 
\par
One striking feature of the TM carbides is that they are almost never found in ideal stoichiometry and often display large deviations from the ideal one-to-one metal-to-carbon ratio.\cite{toth,Williams1966,Williams1971,Williams1988,toth,Schwarz1987} The vacancy concentration in TiC can be as large as 50\%\cite{toth} while still maintaining the rock-salt crystal structure and without any secondary Ti phase. This has, obviously, resulted in many investigations of the properties of these materials when regarding their defect structure, see for example Refs.~\onlinecite{Hugosson2}, \onlinecite{Rasander2017}, \onlinecite{Tan}, \onlinecite{Redinger} \onlinecite{Tsetseris2008}, \onlinecite{Sun2015} and \onlinecite{Ding2012}, as well as references therein. The reasons for the large deviation from ideal stoichiometry have been argued to be due: (i) large number of vacancies formed as a result of the high temperature preparation of TM carbides which leads to an increased configurational entropy stabilising a high vacancy concentration, in connection with subsequent high vacancy migration barriers preventing the system from reaching stoichiometry at low temperatures;\cite{Zhukov1989} (ii) slow diffusion rates at formation temperatures of the TM carbides that inhibits penetration of carbon into the TM lattices;\cite{Klein1980} (iii) that vacancies are energetically stabilised due to the formation of electronic states below the Fermi level which stabilises the vacancies;\cite{Huisman1980} and (iv) interaction and ordering of the vacancies may lead to stabilising carbides with a low C content.\cite{Pickett1986,Ozolins1993,Korzhavyi-2001,Andersson2008} Ordering is especially important in V carbides where several ordered phases with low C content have been found.\cite{toth,Ozolins1993} We note that (i) and (ii) can be regarded as having the origin in the kinetics of the system, while (iii) and (iv) to have the origin in the thermodynamics of the system. However, it is likely that all of these reasons are important for the understanding of the substoichiometry of TM carbides.
\par
In this study we are interested in evaluating the trends in the ability to form C vacancies among the TM carbides using density functional calculations. We will use standard methodology\cite{Zhang-1991,VandeWalle-1993,VandeWalle-2004,Freysoldt-2014} for the evaluation of the C vacancy formation energies in the TM carbides. The present method is valid for low temperatures where the total entropy in the system can be considered to be small. In addition, we do not consider lattice vibrations and the way these vary in the presence of defects. The systems that are of interest are the TM carbides formed by metals of group IV, V and VI, i.e. the strong carbide forming TM. We also include ScC. Since most of these carbides are found in the B1 crystal structure while only MoC and WC are found in the WC-structure, C vacancy formation will be investigated in the B1 structure for all TM, while C vacancy formation in the WC-structure will be analysed for the Group VI TM only. For simplicity, and in order to analyse trends properly, we neglect the fact that Cr does not form a carbide in either the B1 or WC structures.

%%%%%%%%%%%%%%%%%  Formation energy section %%%%%%%%%%%%%
\section{The defect formation energy}
To evaluate the C vacancy formation energy, we use the general theory for calculating defect formation energies.\cite{Zhang-1991,VandeWalle-1993,VandeWalle-2004,Freysoldt-2014} In general for a metallic system of a given supercell size, the defect formation energy is given by
\begin{equation}
E_{f} = \bigl[ E_{D}-E_{H}\bigr] + \sum_{i} n_{i} \bigl(\mu_{i}^{0} + \Delta\mu_{i}\bigr),
\end{equation}
where $E_{D}$ is the total energy of the supercell containing the defect $D$ and $E_{H}$ is the total energy of the ideal host crystal of the same dimensions as the supercell containing the defect. The crystal growth conditions affect the formation energy via the chemical potentials, $\mu_{i}=\mu_{i}^{0}+\Delta\mu_{i}$, of element $i$. These potentials represent the energy cost of exchanging atoms with the chemical reservoir. The chemical potentials are added or subtracted from the formation energy according to the number of atoms ($n_{i}$) of a certain element added or withdrawn from the growth reservoir, where $n_{i}=+1$ if an atom is removed from the system and $n_{i}=-1$ if an atom is added to the system. By convention, the chemical potentials are defined in relation to the standard states of the constituents of a systems. For the metallic elements in the present study, the standard reference states are the hexagonal closed packed (hcp) structure in the case of Sc, Ti, Zr and Hf, and the body centered structure (bcc) in the case of V, Cr, Nb, Mo, Ta and W. For carbon, the reference state is graphite.
\par
The chemical potentials are broken into two parts where $\mu_{i}^{0}$ is the chemical potential of the element of the reference state of the element $i$ and $\Delta\mu_{i}$ is the chemical potential of the element in relation to to its reference state. The growth conditions are reflected in $\Delta\mu_{i}$: A maximally rich growth environment of an element has $\Delta\mu_{i}=0$. As the concentration of element $i$ decreases during crystal formation, $\Delta\mu_{i}$ also decreases and thus becomes increasingly negative. Assuming equilibrium growth conditions, the chemical potentials are restricted to values that maintain a stable compound and do not permit competing phases to exist. For TM carbides the following constraints apply: In order to avoid precipitates of the elemental forms of TM and C, $\mu_{\rm TM}\leq\mu_{\rm TM}^{0}$ and $\mu_{\rm C}\leq\mu_{\rm C}^{0}$. The stability criterion for the carbide phases are
\begin{equation}
\Delta\mu_{\rm TM} + \Delta\mu_{\rm C} \leq \Delta H_{f}({\rm TMC}), 
\end{equation}
where $\Delta H_{f}({\rm TMC})$ is the energy of formation of the TM carbide.
\par
By combining the above equations for the chemical potentials, it is found that the chemical potentials for the TM and C varies over a range that is identical to the energy of formation of the TM carbide, and in particular the vacancy formation energy for a C vacancy in TM carbides is given by
\begin{equation}\label{eq:C-rich}
E_{f} = \bigl[E_{D} - E_{H}\bigr] +\mu_{\rm C}^{0},
\end{equation}
for C-rich (TM-poor) conditions, and by
\begin{equation}\label{eq:TM-rich}
E_{f} = \bigl[E_{D} - E_{H}\bigr] +\mu_{\rm C}^{0} + \Delta H_{f}({\rm TMC}),
\end{equation}
for C-poor (TM-rich) conditions. It is therefore clear that TM carbides with a strong ability to form carbides, i.e. having a large energy of formation, will have a larger difference between the C vacancy formation energy evaluated for C-rich and TM-rich conditions. 
\par

%%%%%%%%%%%%%%% Details of the calculations.
\begin{table}[t]
\caption{\label{tab:reference} Calculated lattice constants of TM carbides in the B1 crystal structure. $a$ is the lattice constant of the TM carbides evaluated from DFT. The data within parenthesis are taken from Vojvodic and Ruberto\cite{Vojvodic} and H{\"a}glund {\it et al.}\cite{Haglund} $a_{\text{exp}}$ is the experimental lattice constant and the values are obtained from Ref.~\onlinecite{toth} unless otherwise specified. Furthermore $\Delta a = a-a_{\text{exp}}$.}
\begin{ruledtabular}
\begin{tabular}{lccc}
Carbide & $a$ (\AA) & $a_{\text{exp}}$ (\AA)& $\Delta a/a$\\%& Energy (eV/f.u.)\\
\hline
\\
 ScC & 4.673 (4.684\cite{Vojvodic},4.524\cite{Haglund})& 4.637\cite{25}& 0.8\%\\%& \\
\\
\hline
\\
 TiC   & 4.337 (4.332\cite{Vojvodic}, 4.064\cite{Haglund}) & 4.327, 4.325\cite{33} & 0.2~\% \\
 ZrC    & 4.710 (4.702\cite{Vojvodic}, 4.509\cite{Haglund}) & 4.698, 4.691\cite{33} & 0.6~\% \\
 HfC    & 4.646 ( - , 4.440\cite{Haglund}) & 4.640 & 0.2~\%\\
\\
\hline
\\
 VC    & 4.155 (4.164\cite{Vojvodic}, 3.789\cite{Haglund}) & 4.166, 4.163\cite{33} & -0.3~\% \\
 NbC  & 4.506 (4.492\cite{Vojvodic}, 4.233\cite{Haglund}) & 4.470, 4.454\cite{33} & 0.8~\%\\
 TaC   & 4.474 (4.479\cite{Vojvodic}, 4.223\cite{Haglund}) & 4.456, 4.453\cite{33} & 0.4~\%\\
\\
\hline
\\
 CrC   & 4.068 ( - , 3.604) & - & -\\
 MoC  & 4.376 (4.450\cite{Vojvodic}, 4.027\cite{Haglund}) & 4.281, 4.270\cite{18} & 2.2~\% \\
 WC    & 4.378 (4.382, 4.059) & 4.220 & 3.7~\%\\
 \end{tabular}
\end{ruledtabular}
\end{table}
%%%%%%%%%%%%%
\begin{table}[t]
\caption{\label{tab:structure-WC} Calculated lattice constants for the TM carbide in the WC structure.}
\begin{ruledtabular}
\begin{tabular}{lcccc}
Carbide & $a$ (\AA) & $a_{\rm exp}$ (\AA)  & $c$ (\AA) & $c_{\rm exp}$ (\AA) \\
\hline
  CrC & 2.704 & - & 2.628 & - \\
 MoC &  2.917 & 2.898\cite{toth} & 2.828 & 2.809\cite{toth} \\
  WC & 2.919 & 2.906\cite{toth} & 2.845 & 2.837\cite{toth}\\
 \end{tabular}
\end{ruledtabular}
\end{table}
%%%%%%%%%%%%%% Details of the calculations %%%%%%%%%%%%%%
\section{Details of the calculations}
The calculations have been performed using density functional theory, where we have used Bl{\"o}chls projector augmented wave (PAW) method\cite{bloechl} and a plane-wave basis set in order to solve the Kohn-Sham equation as implemented in the Vienna {\it ab initio} simulation package.\cite{KresseandFurth,KresseandJoubert}. The calculations have been performed in a scalar relativistic fashion, i.e. neglecting spin-orbit coupling, and also without considering spin-polarisation of the electron density. We have made use of the generalized gradient approximation of Perdew, Burke, and Ernzerhof (PBE) for the exchange and correlation functional.\cite{PBE} The k-points have been set up using the special k-points method of Monkhorst and Pack\cite{MonkhorstandPack} such that the maximal spacing between k-points is 0.1~\AA$^{-1}$. This corresponds to a 26$\times$26$\times$26 k-point mesh for the primitive unit cell of TiC. A plane wave energy cut-off of 800~eV has been used in all calculations and relaxations of the atomic geometry have been performed until the forces acting on the atoms where smaller than 1~meV/{\AA}. 
\par
With these settings we obtain the lattice constants of the B1 and WC structures as shown in Tables~\ref{tab:reference} and \ref{tab:structure-WC}. Overall, the lattice constants are in good agreement with experiments, with a deviation from the experimental references of about 1~\% or smaller. We note that the experimental lattice constants depend slightly on the number of carbon vacancies in the carbide,\cite{Lewin2010} however, considering that the vacancy concentrations for these reference values are rather small, any large deviation from these references is not expected. Also shown in Table~\ref{tab:reference} are data taken from two previous DFT studies of bulk TM carbides: Vojvodic and Ruberto\cite{Vojvodic} have made used of plane-waves and Vanderbilts ultra-soft pseudopotentials\cite{US-PP} within the GGA approximation (PW91)\cite{PW91} while H{\"a}glund et al.\cite{Haglund} made use of a full-potential linearized muffin-tin orbitals (FP-LMTO) method\cite{LMTO} within the local density approximation (LDA)\cite{LDA}. We note that our data are close to the data of Vojvodic and Ruberto, while the data from H{\"a}glund et al. are consistently smaller than ours. Furthermore, the lattice constants obtained by H{\"a}glund et al. are consistently smaller than the experimental values, which is due to the typical overbinding obtained when using the LDA.
\par
To model the isolated defects we have used a 5$\times$5$\times$5 repetition of the primitive B1 and WC-structured unit cells. Both the B1- and WC-structured supercells contain 250 atomic positions and the B1-structured supercells have been used previously to model defects in TiC.\cite{Rasander2017} We note that this size of supercell provides an effective vacancy concentration of 0.8\%. The use of a finite size supercell introduces an error in the defect formation energies since defects in neighbouring cells will have a non-physical interaction with each other. Usually it is possible to use supercells that are large enough to minimise spurious interactions between neighbouring cells, such that the finite size error is small enough for whatever property that is of interest. In a previous study we used a finite size scaling procedure to calculate the C vacancy formation energy in TiC.\cite{Rasander2017} The C vacancy formation energy was evaluated for a set of supercells of increasing size and thereafter interpolated to evaluate the vacancy formation energy of a single C vacancy in an infinite supercell.\cite{Rasander2017} The difference in the C vacancy formation energy between the extrapolated result and the result obtained for the 5$\times$5$\times$5 supercell was found to be 0.12 to 0.17~eV depending on which density functional approximation was used, where the extrapolated value is the lower value. We consider it to be likely that this difference is the same or similar for all TM carbides in our study and since we are  interested in the trends in the vacancy formation energies among the TM carbides we will use the 5$\times$5$\times$5 supercell to evaluate the C vacancy formation energy in the TM carbides.
\par
In order to calculate the C vacancy formation energies in Eqns.~(\ref{eq:C-rich}) and (\ref{eq:TM-rich}) it is required to calculate the chemical potential of carbon in its graphite reference state. This is problematic when using standard density functional approximations since these approximations lack a proper description of the van der Waals interactions between the graphene sheets in graphite.\cite{doi:10.1021/jp106469x} In a recent study it was shown that it is possible to simulate graphite using standard density functionals by evaluating the energy of single sheets of graphene separately using density functional calculations and to add the interatomic van der Waals interaction energy evaluated using the random phase approximation (RPA).\cite{Rasander2017} As in Ref.~\onlinecite{Rasander2017} we use the RPA interlayer interaction energy obtained by Leb{\`e}gue {\it et al.}\cite{Lebegue2010} of $E_{\rm vdW-RPA}^{\rm C} = 48$~meV. By using this convention to calculate the total energy of graphite it is possible to avoid the evaluation of the van der Waals bonding in this system by DFT. Furthermore, it has been shown that using this convention makes graphite and not diamond the ground state when using the local density approximation (LDA),\cite{Rasander2017} which is not the case if LDA is used without the RPA correction.\cite{Janotti2001}

%%%%%%%%%%%%% Density of states plots %%%%%%%%%%
\begin{figure}[t]
\includegraphics[width=8cm]{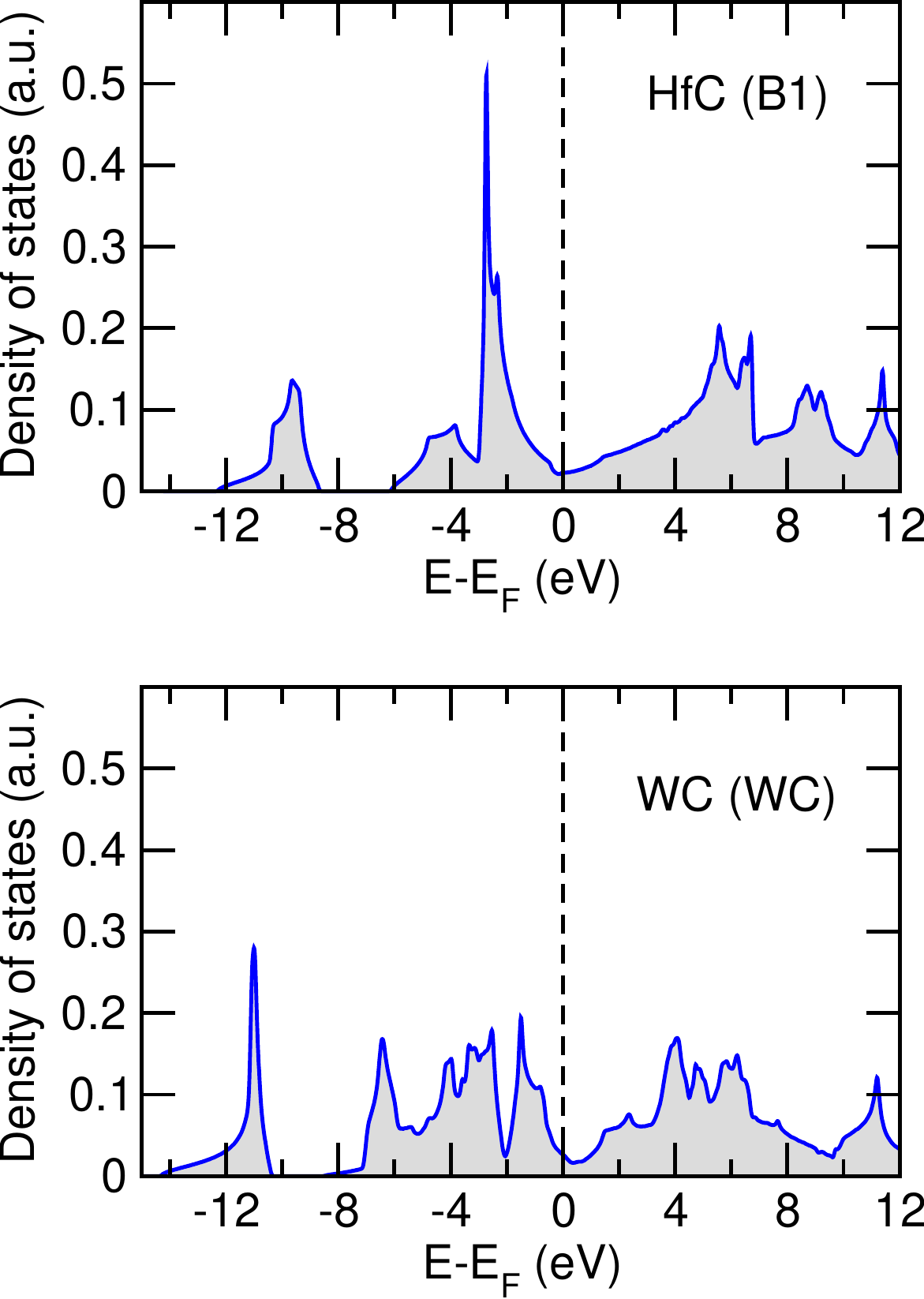}
\caption{\label{fig:DOS} (Color online) Calculated density of states of HfC and WC calculated in the B1 and WC structures respectively. The Fermi level, $E_{F}$, is marked by the dashed vertical line.}
\end{figure}

%%%%%%%%%%%%%%% Vacancy formation energy %%%%%%%%%%%%
\begin{figure*}[t]
\includegraphics[width=14cm]{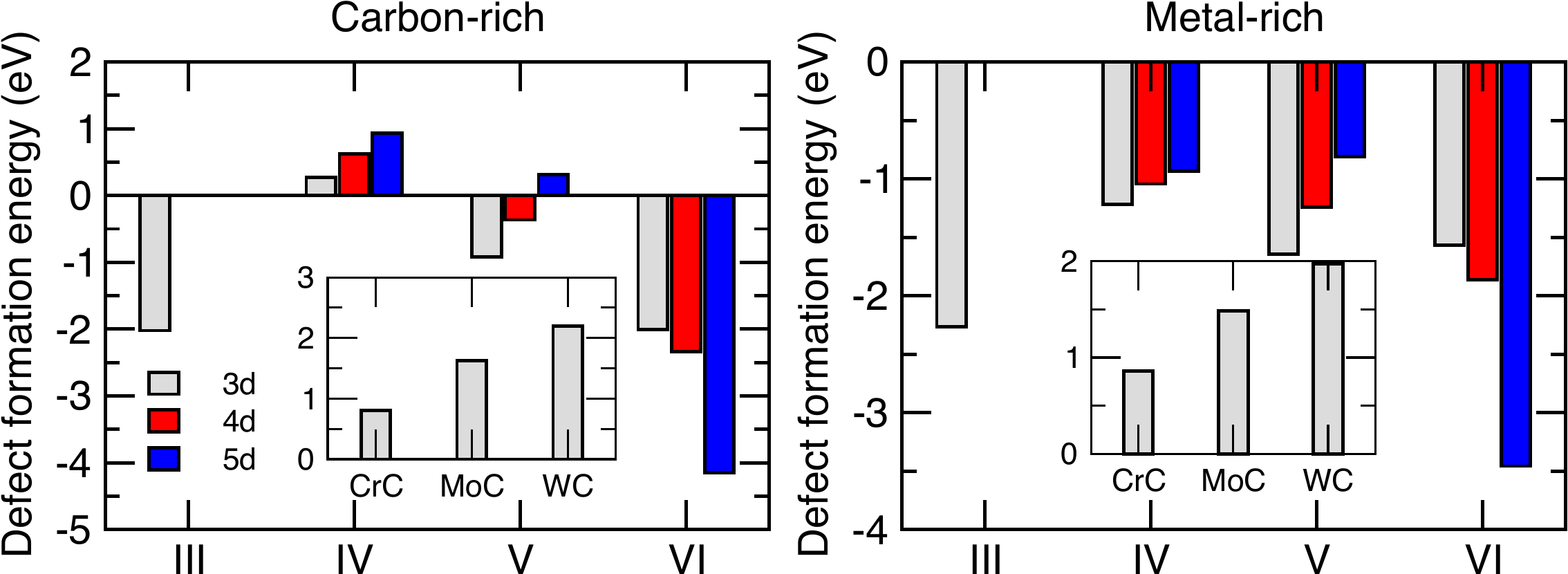}
\caption{\label{fig:formE} (Color online) Calculated carbon vacancy formation energies in TM carbides for carbon-rich and metal-rich conditions. TM carbides with the metal atom from the 3d, 4d and 5d TM series are represented by grey, red and blue bars, respectively. The insets show the calculated carbon vacancy formation energies of TM carbides in the WC crystal structure. All formation energies are given in eV.} %These values are also shown in Tables~\ref{tab:vacancyform} and \ref{tab:vacancyform-WC}.}
\end{figure*}

%%%%%%%%%%%%%%%% Results %%%%%%%%%%%%%%%%%%%
\section{Results}

\subsection{The heat of formation of B1 and WC structured TM carbides}
%%%%%%%%%%%%%%% Heats of formation %%%%%%%%%%%%%%%
\begin{table}[b]
\caption{\label{tab:heatofformation} Calculated heats of formation for TM carbides in the B1 and WC structure types.}
\begin{ruledtabular}
\begin{tabular}{lcc}
Carbide & $\Delta H_{f}^{\rm B1}$ (eV) & $\Delta H_{f}^{\rm WC}$ (eV) \\
\hline
\\
 ScC & -0.24 & - \\
\\
\hline
\\
 TiC & -1.49 & - \\
ZrC & -1.67 & - \\
  HfC & -1.87 & - \\
\\
\hline
\\
  VC & -0.73 & - \\
 NbC & -0.88 & -\\
  TaC & -1.13 & -\\
\\
\hline
\\
  CrC & 0.44 &  0.06\\
 MoC & 0.48 &  -0.14\\
  WC & 0.69 & -0.23\\
 \end{tabular}
\end{ruledtabular}
\end{table}

In Table~\ref{tab:heatofformation}, we show the calculated heat of formation for the TM carbides. By definition, negative values suggest a stable compound, i.e. the compound has a lower energy compared to having the TM and graphite as two separate phases, while positive values suggest unstable compounds. Our calculations show that the TM carbides derived from Group III, IV and V TM are stable while Group VI TM carbides are unstable in the B1 structure. Furthermore, we note that the Group IV TM carbides are the most stable systems, and that for the stable TM carbide phases the stability of a TM carbide increases by moving downwards along a group, such that HfC is more stable than ZrC and so on.
\par
The behaviour of the heats of formation is an effect of the electronic structure of the TM carbides in the B1 structure. As an illustration, the density of states of HfC in the B1 structure is shown in Fig.~\ref{fig:DOS}. It is well known that the group IV TM carbides have the Fermi-level positioned in a valley in the DOS which separates bonding from anti-bonding states.\cite{Haglund,Rasander,Grechnev} The group IV TM carbides are therefore maximally bonded while the extra electron in the case of the group V TM carbides occupies anti-bonding states and, as a result, the group V TM carbides have a lower heat of formation since the extra electron derived from the Group V TM occupies antibonding states at higher energies which increases the total energy of the TM carbide. In the case of Group VI TM elements the two additional electrons per TM populate even more antibonding states which increases the total energy even further, with a very weak ability of forming B1 structured TM carbides as a result, as is clearly seen in Table~\ref{tab:heatofformation}. For ScC, we note that the lower electron count in ScC compared to TiC makes sure that not all of the binding states are occupied and the total energy in ScC is therefore higher than in TiC, with a weaker ability of forming a B1 structured TM carbide than TiC.
\par
The heats of formation of TM carbides in the WC structure is also shown in Table~\ref{tab:heatofformation}. It is clear that for the Group VI TM it is much more favourable to form a carbide in the WC structure than in the B1 structure, even though the heat of formation of CrC still suggest that the system is unstable. As was the case for the B1 structured carbides, the energetic stability of the WC structured carbides increases when moving downwards along the Group VI TM elements. We also note that the heat of formation for the WC structured carbides are smaller than in the case of the stable B1 structured carbides. 
\par
In Fig.~\ref{fig:DOS} we also show the density of states of WC in the WC structure. We note that the same division between bonding and antibonding states can be made for this system,\cite{Hugosson2003} with the difference that the maximally bonded systems are now the TM carbides that are formed using the Group VI TM elements Cr, Mo and W.

\subsection{The C vacancy formation energy of B1 structured TM carbides}
Before analysing the C vacancy formation energies we need to mention that in a normal crystal, defects are stabilised due to the increased entropy that is obtained by the introduction of the defect. This is called entropic stabilisation, i.e. defects increase the energy of the system, but the increased entropy that accompanies the introduction of the defects makes them stable. Entropic stabilisation suggests that only a limited number of defects should be present at any finite temperature, and the defect concentration increases as the temperature is raised. It is also possible to imagine energetic stabilisation of defects, where the formation of defects lowers the energy of the system and therefore the defects are stabilised even without the increased entropy. Both these types of stabilisation of C vacancies can be found in the TM carbides.
\par
In Fig.~\ref{fig:formE}, we present the calculated C vacancy formation energies for the TM carbides in the B1 and WC crystal structures. For the B1 structured compounds we find that the highest C vacancy formation energy is found for HfC and the lowest formation energy is found for WC for C-rich conditions. In fact, ScC, VC, NbC, CrC, MoC and WC all have negative vacancy formation energies which suggests that carbon vacancies will be formed spontaneously in these TM carbides, even at conditions rich in carbon. The vacancies in these carbides are therefore energetically stabilised for C-rich conditions. 
\par
For C-rich conditions, TiC, ZrC, HfC and TaC have positive C vacancy formation energies and, therefore, it will cost energy to form carbon vacancies in these systems. At equilibrium conditions, i.e. excluding non-equilibrium growth processes, C vacancies in these four carbides are therefore entropically stabilised in C-rich conditions. We note that these TM carbides are also the most stable systems according to their heats of formation as shown in Table~\ref{tab:heatofformation}.
\par
For TM-rich conditions we find that all B1 carbides have a negative C vacancy formation energy and therefore there will be a spontaneous driving force for the creation of C vacancies for such conditions in these systems. This feature of the C vacancy formation energy is derived from the ability of TM carbides to form strong bonds between metal and carbon atoms: It is more preferable for a TM to bind in a carbide phase even if that implies the formation of C vacancies. That there is a dramatic shift from positive to negative C vacancy formation energies as the growth conditions varies from C-rich to TM-rich in TiC, ZrC, HfC and TaC is due to the large heats of formation of these compounds, as reflected in Eqns.~(\ref{eq:C-rich}) and (\ref{eq:TM-rich}). We also note that the C vacancy formation energies for TiC, ZrC, HfC and TaC are highly skewed toward negative values since the vacancy formation energy for TM-rich conditions are more negative than the vacancy formation energy is positive for C-rich conditions. This means that for a large range of intermediate conditions between TM-rich and C-rich extremes the C vacancy formation energies for these systems will be negative and C vacancies will be energetically stabilised. 
\par
Furthermore, we find that within each group the vacancy formation energy increases when moving downwards, i.e. substituting the 3d TM for a 4d TM and so on, except for the Group VI TM carbides where the opposite trend is found. This is in line with the experimental phase diagrams of these compounds.\cite{toth} In addition, when following along a period, the maximum in the vacancy formation energy occurs for the Group IV TM carbides, for each of the 3d, 4d and 5d TM carbide series. We note that TaC has a positive vacancy formation energy at C-rich conditions while the other Group V TM carbides have negative formation energies irrespective of the conditions.
\par

\subsection{The C vacancy formation energy of WC structured TM carbides}

In general it is relatively easy for C vacancies to be formed in the B1 crystal structure, especially for TM-rich conditions. If we move on to the WC structured systems, we find that the C vacancy formation energies are all positive for both C-rich and TM-rich conditions. Furthermore, the vacancy formation energy are generally larger in the WC structure than for the B1 structured TM carbides, which means that C vacancies are less likely to be formed in TM carbides crystallised in the WC structure compared to TM carbides formed in the B1 structure. This results agrees well with the experimental fact that the WC structure is only found at the one-to-one metal to carbon ratio.\cite{toth} 
\par
That the C vacancy formation energy in the WC structures is larger than in the B1 crystal structures is interesting when considering that the heats of formation of the WC structured carbides are small related to the stable B1 structured carbides. This suggests that even though the heat of formation, which is a measure of the bond strength in the carbide, is small the WC structured TM carbides are less able to stabilise vacancies in the lattice compared to TM carbides formed in the B1 structure.

\begin{table*}[t]
\caption{\label{tab:relaxations} First ($d_{n}^{(1)}$), second ($d_{n}^{(2)}$) and third ($d_{n}^{(3)}$) nearest neighbour distances from a C vacancy site in the B1 structure. The values given in parenthesis are the corresponding distances in the pristine TM carbide. $\Delta d_{n}^{(1)}$, $\Delta d_{n}^{(2)}$ and $\Delta d_{n}^{(3)}$ are the differences between the first, second and third nearest neighbour distances from a defect site in relation to the same distances in the pristine TM carbide.}
\begin{ruledtabular}
\begin{tabular}{lccccccc}
Carbide  & $d_{n}^{(1)}$ (\AA) & $d_{n}^{(2)}$ (\AA)& $d_{n}^{(3)}$ (\AA) &  $\Delta d_{n}^{(1)}$ & $\Delta d_{n}^{(2)}$ & $\Delta d_{n}^{(3)}$ \\
\hline
\\
 ScC  & 2.469 (2.337) & 3.256 (3.305) & 4.035 (4.047) &  0.132 (5.6~\%) & -0.049 (-1.5~\%) & -0.012 (-0.3~\%)\\
\\
\hline
\\
TiC   & 2.269 (2.168) & 3.049 (3.067) & 3.742 (3.756) & 0.101 (4.7~\%) & -0.018 (-0.6~\%) & -0.014 (-0.4~\%)\\
 ZrC  & 2.456 (2.355) & 3.312 (3.330) & 4.069 (4.079) & 0.102 (4.3~\%) & -0.019 (-0.6~\%) & -0.010 (-0.2~\%)\\
 HfC   & 2.419 (2.323) & 3.267 (3.285) & 4.013 (4.024) & 0.096 (4.1~\%) & -0.018 (-0.5~\%) & -0.010 (-0.2~\%)\\
\\
\hline
\\
 VC     & 2.214 (2.077) & 2.916 (2.938) & 3.544 (3.598) & 0.137 (6.6~\%) & -0.022 (-0.7~\%) & -0.054 (-1.5~\%)\\
 NbC   & 2.353 (2.253) & 3.173 (3.187) & 3.861 (3.902) & 0.100 (4.4~\%) & -0.013 (-0.4~\%) & -0.042 (-1.1~\%)\\
 TaC     &  2.291 (2.237) & 3.160 (3.164) & 3.840 (3.875) & 0.053 (2.4~\%) & -0.004 (-0.1~\%) & -0.035 (-0.9~\%) \\
\\
\hline
\\
  CrC  & 2.285 (2.034) & 2.851 (2.877) & 3.308 (3.523) & 0.251 (12.3~\%) & -0.025 (-0.9~\%) & -0.216 (-6.1~\%)\\
 MoC  & 2.344 (2.188) & 3.013 (3.094) & 3.905 (3.789) & 0.157 (7.2~\%) & -0.081  (-2.6~\%) & 0.116 (3.1~\%)\\
  WC  & 1.981 (2.189) & 3.126 (3.096) & 3.649 (3.791) & -0.208 (9.5~\%) & 0.030 (1.0~\%) & -0.143 (-3.8~\%)\\
 \end{tabular}
\end{ruledtabular}
\end{table*}

\begin{table*}[t]
\caption{\label{tab:relaxations-WC} First ($d_{n}^{(1)}$), second ($d_{n}^{(2)}$) and third ($d_{n}^{(3)}$) nearest neighbour distances from a C vacancy site in the WC structure. The values given in parenthesis are the corresponding distances in the pristine TM carbide. $\Delta d_{n}^{(1)}$, $\Delta d_{n}^{(2)}$ and $\Delta d_{n}^{(3)}$ are the differences between the first, second and third nearest neighbour distances from a defect site in relation to the same distances in the pristine TM carbide.}
\begin{ruledtabular}
\begin{tabular}{lccccccccc}
Carbide & $d_{n}^{(1)}$ (\AA) & $d_{n}^{(2c)}$ (\AA)& $d_{n}^{(2a)}$ (\AA) &  $d_{n}^{(3)}$ (\AA) & $\Delta d_{n}^{(1)}$ & $\Delta d_{n}^{(2c)}$ & $\Delta d_{n}^{(2a)}$ & $\Delta d_{n}^{(3)}$ \\
\hline
  CrC & 2.074 (2.039) & 2.569 (2.622)  & 2.664 (2.704) & 3.378 (3.386) & 0.035 (1.7~\%) & -0.053 (-2.0~\%) & -0.040 (-1.5~\%) & -0.009 (-0.3~\%)\\
 MoC & 2.221 (2.199) & 2.809 (2.828) & 2.882 (2.917) & 3.642 (3.653) & 0.022 (1.0~\%) & -0.019 (-0.7~\%) & -0.035 (-1.2~\%) & -0.011 (-0.3~\%)\\
  WC & 2.226 (2.205) & 2.834 (2.845) & 2.888 (2.919) & 3.647 (3.658) & 0.021 (1.0~\%) & -0.011 (-0.4~\%) & -0.031 (-1.1~\%) & -0.011 (-0.3~\%)\\
 \end{tabular}
\end{ruledtabular}
\end{table*}

\subsection{Local relaxations due to C vacancies in B1 and WC structured TM carbides}

In Table~\ref{tab:relaxations} we show the resulting structure close to the vacancy position in the B1 structure. For this structure, the nearest neighbour position, $d_{n}^{(1)}$, in a TM carbide is a TM to C binding distance, while the second nearest neighbour distance, $d_{n}^{(2)}$, is the smallest TM to TM or C to C distance in the crystal. $d_{n}^{(3)}$ is again a TM to C distance. In the ideal structure, $d_{n}^{(1)}=a/2$, $d_{n}^{(2)}=a/\sqrt{2}$ and $d_{n}^{(3)} = \sqrt{3}a/2$. Since we are interested in the structural relaxations due to the presence of C vacancies, $d_{n}^{(2)}$ measures the distance between the C vacancy and its nearest C neighbours.
\par
In Table~\ref{tab:relaxations} the three nearest neighbour distances are given for pristine TM carbides as well as the corresponding distances from a C vacancy site to its nearest neighbours. We find that the TM atoms at the atomic shell closest to the C vacancy moves away from the vacancy. The size of this relaxation varies slightly depending on the TM carbide, with changes ranging from 4.1~\% in the case of HfC to 12.3~\% in the case of CrC. The next atomic shell around the vacancy is relaxing towards the vacancy in all systems except for WC. The relaxations of this shell are smaller with a maximum change of 2.6~\% for MoC. The third atomic shell also relaxes towards the vacancy site. The size of the relaxations in this shell is of the same order or even larger than for the second atomic shell. In general, the relaxations in the B1 structured carbides due to the presence of the C vacancy are rather long range and in order to accommodate the local relaxations the supercell that is used needs to be large enough to allow for enough atomic shells to relax.\cite{Rasander2017} 
\par
In the WC structure, the nearest neighbour distance is given by $d_{n}^{(1)} = \sqrt{a^2/3+c^2/4}$. This is the distance between a C atom to its nearest neighbouring TM atom. The second nearest neighbour is the distance between two C atoms. However, in the WC structure there are two different C to C binding distances given by $d_{n}^{(2c)}=c$ and $d_{n}^{(2a)}=a$ that are very similar. In fact, the difference between these distances is small and even identical if $a=c$, however, as is the case for the TM carbides studied here $c<a$, as is shown in Table~\ref{tab:relaxations-WC}. It is clear from Table~\ref{tab:relaxations-WC} that the relaxations around the C vacancy are much smaller than the corresponding relaxations in the B1 structured carbides. In the WC structure, the first shell moves away from the vacancy, while the second and third shells moves towards the vacancy, i.e., the same relaxation pattern as in the B1 structure with the difference that the relaxations are smaller in the WC structured carbides. 
\par
Table~\ref{tab:relaxation-WC} shows the relaxation energy, $E_{r}$, which is evaluated as the difference in the completely relaxed structure and the structure containing the C vacancy but with all other atoms at their pristine crystal positions. Since the relaxations always work towards stabilising the vacancy, the relaxation energy is negative. We note that without relaxations the C vacancy formation energies will be too high. The largest relaxation energies are found for B1 structured CrC, MoC and WC, while the smallest relaxation energies are found for the same TM carbides but now in the WC structure. In general, the relaxation energy correlates well with the C vacancy formation energy; if the relaxation energy is larger then the vacancy formation energy is smaller.

\begin{table}[t]
\caption{\label{tab:relaxation-WC} Calculated relaxation energies for TM carbides in the B1 and WC structures.}
\begin{ruledtabular}
\begin{tabular}{lcc}
Carbide &  $E_{r}^{\rm B1}$ (eV) & $E_{r}^{\rm WC}$ (eV)\\
\hline
\\
 ScC & -0.67 & - \\
\\
\hline
\\
 TiC & -0.74  & -\\
  ZrC & -0.79 & - \\
 HfC & -0.75 & - \\
\\
\hline
\\
 VC & -0.79 & - \\
 NbC & -0.42 & - \\
 TaC & -0.20 & - \\
\\
\hline
\\
 CrC & -1.05 &  -0.22\\
 MoC & -0.99 & -0.15 \\
 WC & -2.74 & -0.13\\
 \end{tabular}
\end{ruledtabular}
\end{table}

\par

\section{Summary and conclusions}
We have performed density functional calculations in order to investigate the formation of C vacancies in binary TM carbides. In the case of B1-structured TiC, ZrC, HfC and TaC, we find that the ability to form C vacancies depends strongly on the growth conditions: For C-rich conditions, we find that the vacancy formation energy is positive and any thermodynamic stabilisation of vacancies must therefore be an entropic effect. However, vacancies in these systems may also be present due to kinetic effects. For the remaining B1-structured TM carbides investigated here, the vacancy formation energy under C-rich conditions is negative, which means that C vacancies are energetically stabilised in the carbide phase. For TM-rich conditions, we find that the C vacancy formation energy is negative for all B1-structured TM carbides. That the vacancy formation energy is negative for conditions rich in the TM for TiC, ZrC, HfC and TaC is attributed to the strength of the TM-C bond in these systems, since it is more favourable to form a carbide phase with C vacancies than to form a TM carbide with the TM as a secondary phase.
\par
We can therefore conclude that for B1-structured TM carbides, C vacancies are energetically favourable, depending slightly on the TM as well as the growth conditions. This result is in excellent agreement with the experimental observation that large vacancy concentrations are often observed in these TM carbide systems. The exceptions are TiC, ZrC, HfC and TaC which for C-rich conditions all have a positive C vacancy formation energy. That large vacancy concentrations are still found experimentally for these systems is likely due to high activation energy barriers for C diffusion, such that C vacancies present as a results of the high temperature growth process used in these systems are frozen inside the carbide phase, or simply that true C-rich conditions are not obtained in the growth process. 
\par
For the Group VI TM carbides in the WC structure, we find a positive C vacancy formation energy irrespective of the TM as well as the growth conditions. The vacancy formation energy is actually larger for both WC-structured MoC and WC than any C vacancy formation energy for TM carbides in the B1-structure. This is likely the reason for why WC-structured carbides are found only for a small range close to the one-to-one metal-to-carbon ratio.

\section{Acknowledgements}
Financial support from the Swedish e-science Research Centre (SeRC), Vetenskapsr{\aa}det (grant numbers VR 2015-04608 and VR 2016-05980), and Swedish Energy Agency (grant number STEM P40147-1) is acknowledged. The computations were performed on resources provided by the Swedish National Infrastructure for Computing (SNIC) at the National Supercomputer Center (NSC), Link{\"o}ping University, the PDC Centre for High Performance Computing (PDC-HPC), KTH, and the High Performance Computing Center North (HPC2N), Ume{\aa} University.

\bibliography{carbides}
\end{document}